\documentclass[11pt,twoside]{article}

\usepackage{asp2006}
\usepackage{epsf}
\usepackage{psfig}
\usepackage{lscape}
\usepackage{natbib}
\usepackage{graphicx}
\begin{document}
\title{Hydrodynamic Simulations of M51 and the interaction with NGC 5195}   
\author{C. L. Dobbs\altaffilmark{1}, C. Theis\altaffilmark{2}, J. E. Pringle\altaffilmark{3}, and M. R. Bate\altaffilmark{1}}   
\altaffiltext{1}{School of Physics, University of Exeter, Exeter, EX4 4QL}
\altaffiltext{2}{Institute of Astronomy, University of Vienna, T\"urkenschanzstr.\ 17, 1180, Austria}
\altaffiltext{3}{Institute of Astronomy, Madingley Road, Cambridge, CB3 0HA}

\begin{abstract} 
The M51 galaxy is one of best examples of grand design structure in galaxies. The most likely origin of the spiral arms in M51 is the ongoing interaction with NGC 5195. Here we report recent calculations which model the orbit of M51 and NGC 5195. These calculations, which for the first time focus on the gas dynamics, are able to reproduce even detailed features in M51, as well as demonstrate the nature of the spiral structure of M51.
\end{abstract}


\section{Calculations}
We use a Smoothed Particle Hydrodynamics (SPH) method to model M51 and NGC 5195. We take the orbit from \citet{theis2003}, who performed many realisations using a combined N-body and genetic algorithm code, comparing the resulting velocity and density fields to the data of \citet{rots1990}. Their best-fit calculation provides the relative positions and velocities of the two galaxies at the start of our model. We set up the M51 galaxy using the mkkd95 program \citep{kuij1995}, and include a stellar and gaseous disc, and a live halo and bulge. With a total of 1.1 million particles,  this gives a gas mass resolution of 180 M$_{\odot}$. The NGC 5195 galaxy is modelled by a point mass. 
The gas in M51 is isothermal (10$^4$ K) and represents 1\% of the disc mass, chosen deliberately low to avoid stellar collapse which would halt the calculation.

\section{Results}
In Figure~1, we show the time evolution of the M51 galaxy. The galaxy undergoes a transition from a flocculent to a grand design galaxy (in isolation the galaxy is a multi-armed spiral). The 180 and 300 Myr frames indicate that the spiral structure winds up over time, and consequently there is no single pattern speed. The 300 Myr time frame corresponds closest with the current day, at which point the overall shape of the spiral arms, and location of the companion agree well with the observed structure. After 370 Myr, M51 and NGC 5195 start to merge (not shown, but see \citet{dobbs2009}). 

In Fig.~2 we compare the spiral structure in our simulation at a time of 300 Myr with an HST image of M51. As well as capturing the overall shape of the spiral 
arms, we also produce detailed features; a kink in one spiral arm (A), spiral arm bifurcations (B) and 
inter-arm gas (C). The kink occurs where an inner spiral arm, which was induced during the first passage of NGC 5195, meets an outer arm which is due to the second, more recent passage.
Spiral arm bifurcations and branches (and interarm gas) are the result of material shearing away from the spiral arms. A feature not seen in the actual M51 however, is a large amount of gas above the top arm. Intriguingly though, in both simulations and observations there is gas extending from the companion (D), and it is possible this gas extends further but is just not detected away from NGC 5195.

It is evident from Fig.~1 that the spiral pattern winds up over time and in Fig.~3, we plot the pattern speed of the gaseous and stellar spiral arms versus radius. The pattern speed is calculated from the positions of the spiral arms at 180 and 300 Myr, which in turn are determined by fitting Gaussians to the azimuthal density profiles. We see from Fig.~3 that the pattern speed decreases with radius, thus the spiral pattern is not in a quasi-steady state. The arms behave similarly to kinematic density waves, which have a pattern speed of $\Omega_p=\Omega-\kappa/2$ (see also \citealt{oh2008}). The pattern in our simulations actually winds up slightly slower than this, due to self gravity.  However there is only limited 
radially inward propagation of the spiral arms due to self gravity, thus 
they are not actually waves, rather spiral density patterns.

\acknowledgements 
The calculations reported here were performed using the University of
Exeter's SGI Altix ICE 8200 supercomputer.

\begin{figure}
\centerline{
\includegraphics[scale=0.24,viewport=0 0 510 450,clip]{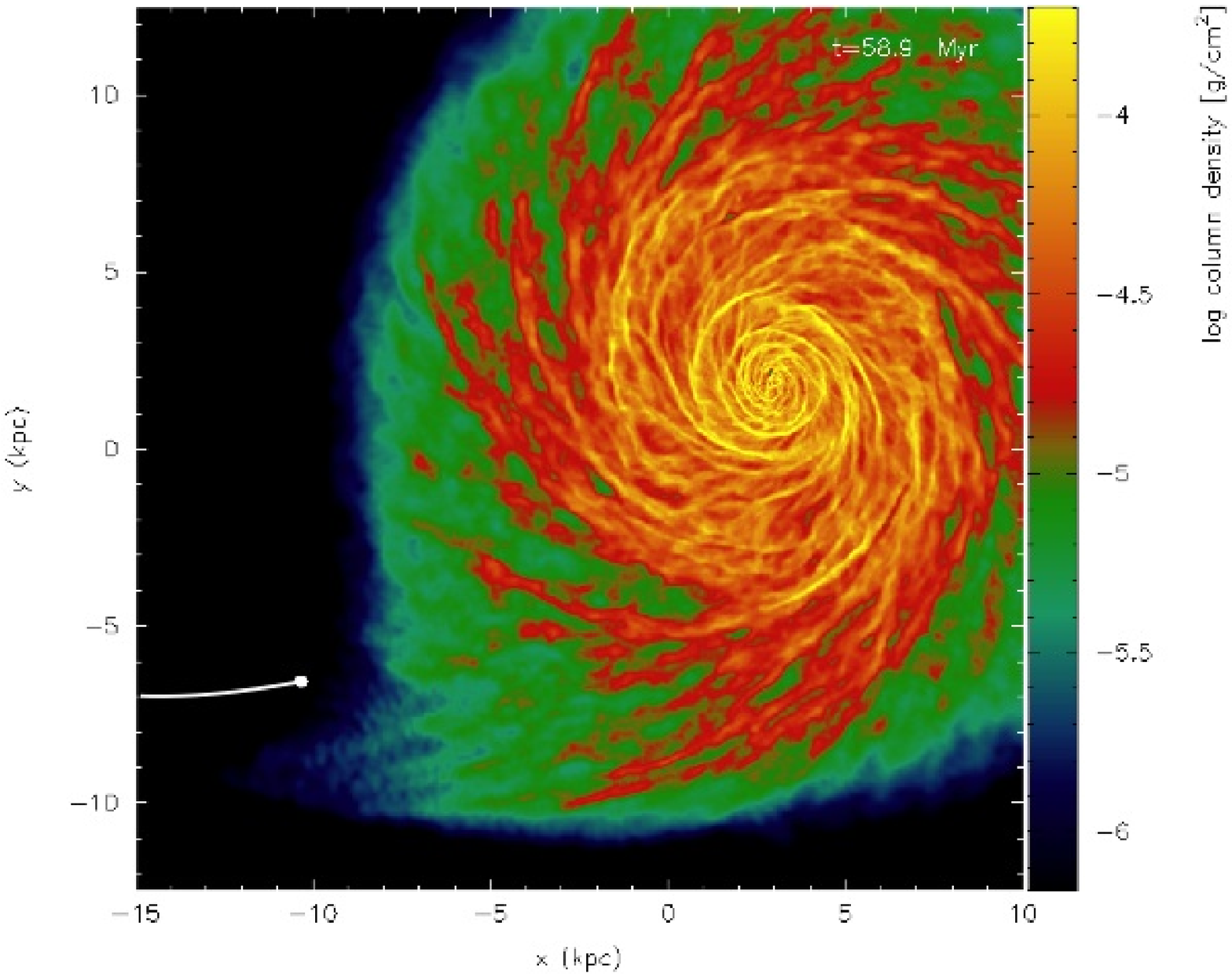}
\includegraphics[scale=0.25,viewport=40 10 510 460,clip]{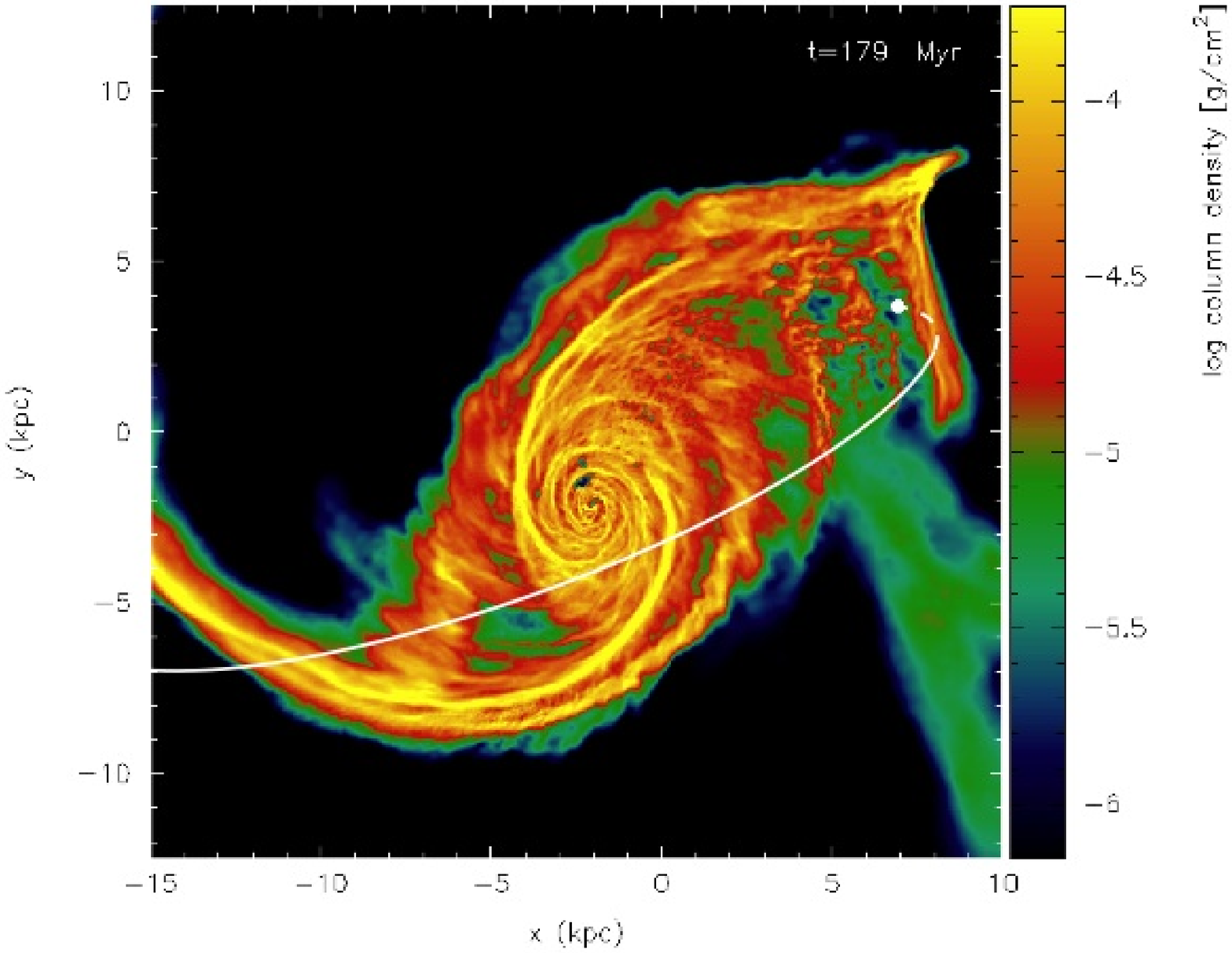}
\includegraphics[scale=0.24,viewport=40 0 550 450,clip]{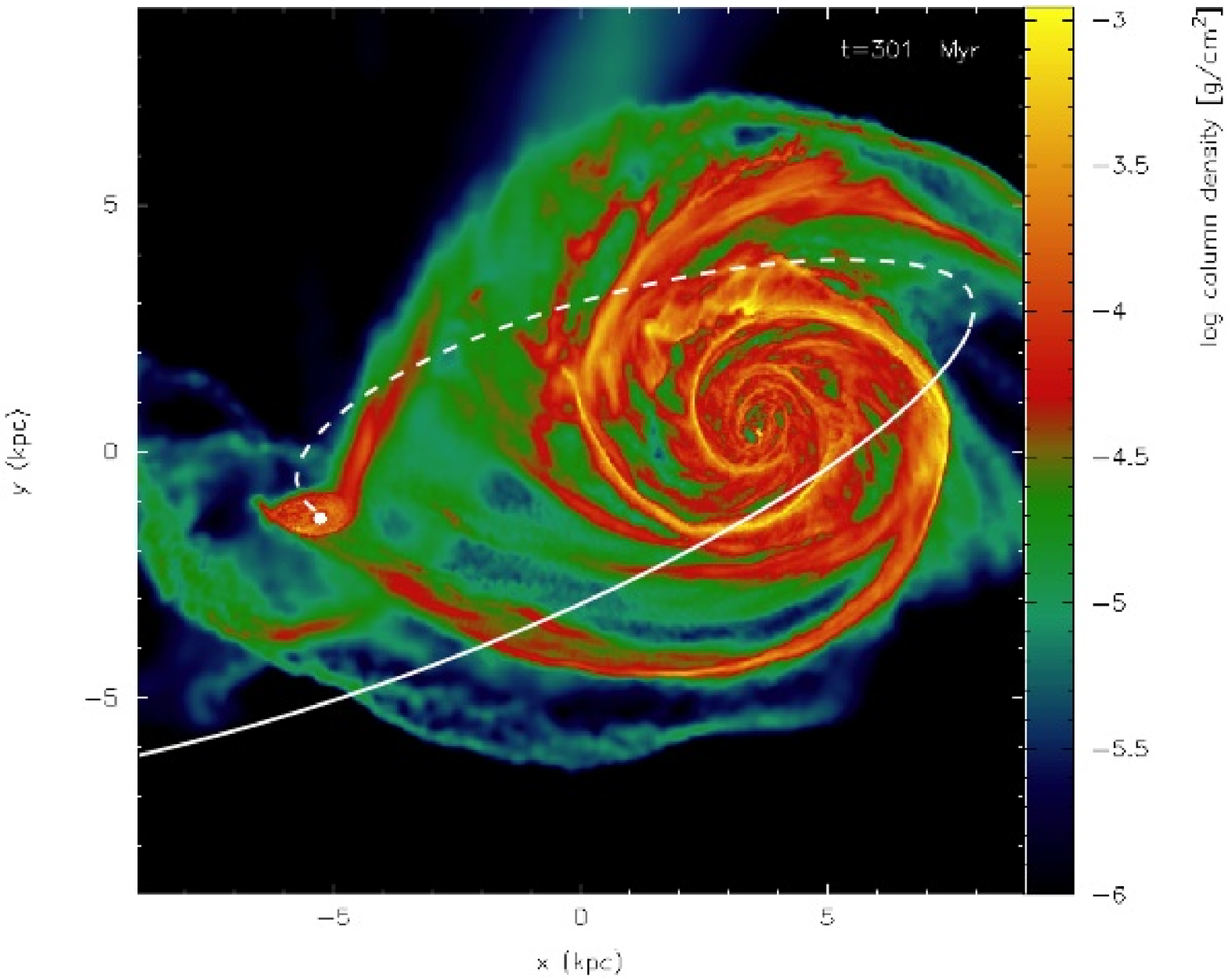}}
\caption{The time evolution of the disc is shown at times of 60, 180 and 300 Myr (the panels show the gas column density).  The M51 galaxy undergoes a transition from a flocculent to a 
grand design galaxy, and will eventually merge with NGC 5195. The 
orbit of NGC 5195 is shown on each panel, and the white dot 
indicates the location of NGC 5195.}
\end{figure}

\begin{figure}
\centerline{
\includegraphics[height=1.9in,viewport=40 40 600 440]{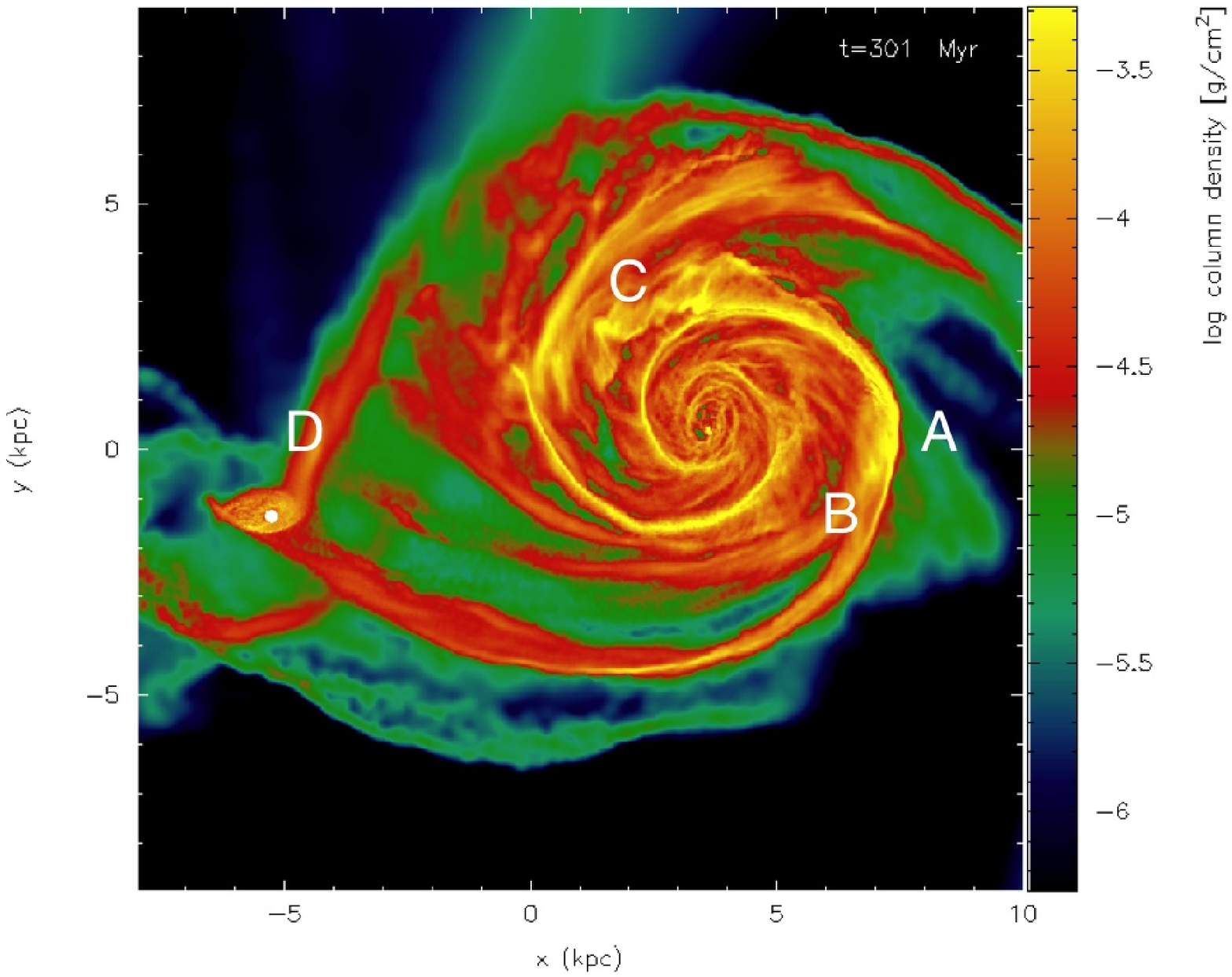}
\includegraphics[height=1.45in,viewport=40 0 550 400]{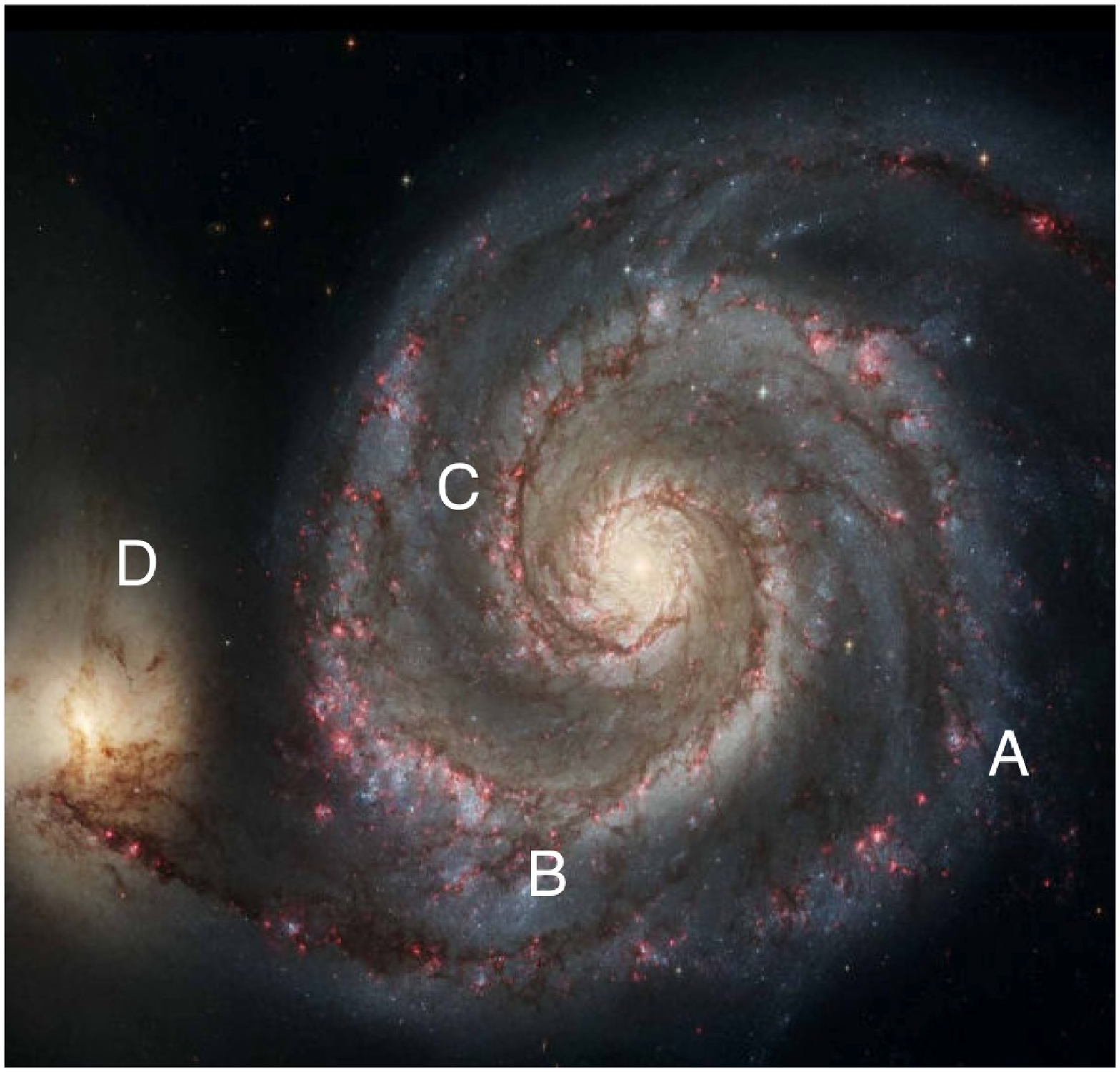}}
\vspace{10pt}
\caption{The gas column density is shown after 300 Myr (left), corresponding to the present day. This figure is displayed alongside a Hubble image of M51, courtesy of S. Beckwith (STScI), and The Hubble 
Heritage Team (STScI/AURA). The simulations capture the shape of the spiral arms very well, and even detailed features such as a kink in one spiral arm (A), spiral arm 
bifurcations (B) and inter-arm gas (C).}
\end{figure}

\begin{figure}
\centerline{
\includegraphics[scale=0.5]{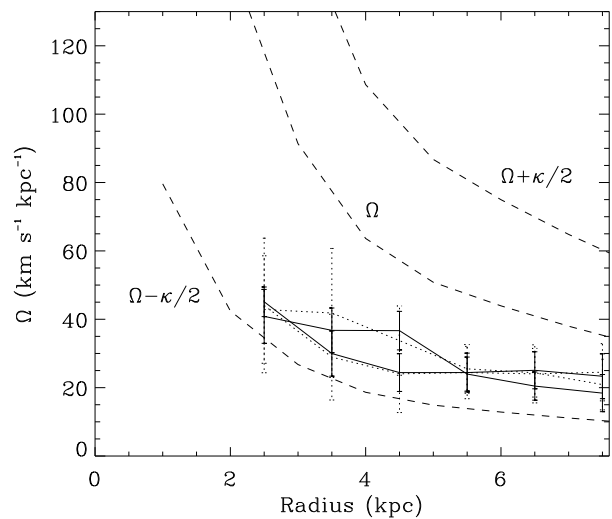}}
\caption{The pattern speed is plotted for the gas (solid) and stellar (dotted) spiral arms, with error bars. There is no single pattern speed rather the pattern speed decreases with radius. This implies the spiral is not a quasi steady spiral density pattern. Meidt et.al. 2008 applied the 
Tremaine-Weinberg method to M51 and found $\Omega_p$=51 kpc km s$^{-1}$ between $r=2$ and $r=4$ kpc, and $\Omega_p$=23 kpc km s$^{-1}$ between $r=4$ and $r=5$ kpc.}
\end{figure}


\end{document}